\documentclass{optica-article}

\journal{optcon}

\articletype{Research Article}

\usepackage{lineno}

\usepackage{bm}
\newcommand{\floor}[1]{\lfloor #1 \rfloor}
\newcommand{\ceil}[1]{\lceil #1 \rceil}

\begin{document}

\title{Interpolated inverse discrete wavelet transforms in additive and non-additive spectral background corrections}

\author{Teemu Härkönen\authormark{1,*
} and Erik Vartiainen\authormark{1}}

\address{\authormark{1} School of Engineering Science, LUT University, Yliopistonkatu 34, Lappeenranta, FI-53850, Finland}

\email{\authormark{*}teemu.harkonen@lut.fi} 



\begin{abstract*}
We demonstrate the applicability of using interpolated inverse discrete wavelet transforms as a general tool for modeling additive or multiplicative background or error signals in spectra.
Additionally, we propose an unsupervised way of estimating the optimal wavelet basis along with the model parameters.
We apply the method to experimental Raman spectra of phthalocyanine blue, aniline black, naphthol red, pigment yellow 150, and pigment red 264 pigments to remove their additive background and to CARS spectra of adenosine phosphate, fructose, glucose, and sucrose to remove their multiplicative background signals.
\end{abstract*}

\section{Introduction}
Raman spectroscopy is a powerful technique for molecule identification and analysis.
As Raman spectroscopy and its variations, such as coherent anti-Stokes Raman scattering (CARS), surface-enhanced Raman spectroscopy (SERS), or surface-enhanced resonance Raman spectroscopy (SERRS), are noninvasive and nondestructive, they are particularly suitable for biomedical applications amongst many others \cite{McNay:2011, Li:2020}.

However, the spectra are often accompanied by some form of a background signal, obfuscating the Raman signal arising from molecular vibrations.
Approaches to correct the background include asymmetric least squares \cite{Boelens:2005, He:2014}, polynomial fits \cite{Gan:2006}, and wavelet-based methods \cite{Galloway:09, Chi:2019}.
For a more extensive review, see \cite{Liland:2010}.

In this study, we concentrate on interpolated discrete wavelet transforms previously considered in \cite{Harkonen:2020} as an extension of the wavelet prism method \cite{Kan:16} to correct CARS multiplicative errors.
We demonstrate the interpolation of wavelet reconstruction levels as a viable way of modeling background processes for general spectra and propose a criterion for automatically choosing the wavelet basis and the associated model parameters.
The model is parameterized with only one or two parameters, enabling applications where high-dimensional parameters could be infeasible such as in Bayesian inference, see for example \cite{Moores:2016, Harkonen:2020}

The rest of the manuscript is structured as follows.
We formulate the interpolated wavelet transform models for additive and multiplicative backgrounds and the optimization scheme in Section \ref{sec:model}.
Experimental details are documented in Section \ref{sec:experimental} and we present results for the experimental Raman and CARS spectra in Section \ref{sec:results}.
We close out the study with conclusions in Section \ref{sec:conclusions}.
\section{Interpolated wavelet background model}
\label{sec:model}
We model a measured spectrum with an additive background $B(\nu_k, c, p, \mathcal{M})$ as
\begin{align}
    y(\nu_k) = f(\nu_k) + B(\nu_k, c, p, \mathcal{M}),
    \label{eq:dataModel}
\end{align}
where $\nu_k$ denotes the measurement location, $c$ a constant offset of the background, $p$ the interpolated wavelet detail level, and $\mathcal{M}$ a chosen wavelet basis which can be considered as a model selection problem.
We model the background as in \cite{Harkonen:2020}, with an interpolation between the discrete wavelet reconstruction levels $D_j(\nu)$ given as
\begin{align}
    B(\nu_k, c, p, \mathcal{M}) = c + \sum\limits_{j = {\ceil{p+1}}}^J D_j(\nu, \mathcal{M}) + (1 - \beta)D_{\ceil{p}}(\nu, \mathcal{M}),
    \label{eq:additive_background}
\end{align}
where $J$ is the maximum chosen wavelet decomposition level, $p \in [1,J]$, $\beta = p - \floor{p}$, and the DWT is computed from the data $y(\bm\nu) = ( y(\nu_1), \dots, y(\nu_K))^T$ using the basis $\mathcal{M}_i$. 
For the non-interpolated approach, see \cite{Kan:16}.

For multiplicative spectra, such as in the case of CARS, the measurement spectrum can be modeled as
\begin{align}
    z(\nu_k) = S( \nu_k )\varepsilon_{\rm{m}}(\nu_k; c, p, \mathcal{M}),
    \label{eq:multiplicativeDataModel}
\end{align}
where $z(\nu_k)$ is a measurement at $\nu_k$, $S(\nu_k)$ denotes the error-free spectrum and $\varepsilon_{\rm{m}}(\nu_k; c, p)$ is an modulating background or error function.
The expression in Eq.\eqref{eq:multiplicativeDataModel} can be made additive by taking the logarithm:
\begin{align}
    \log z(\nu_k) = \log S( \nu_k ) + \log \varepsilon_{\rm{m}}(\nu_k; c, p, \mathcal{M}),
    \label{eq:logMultiplicativeDataModel}
\end{align}
which enables us to model the modulating error function as above:
\begin{align}
    \log\varepsilon_{\rm{m}}(\nu; c, p, \mathcal{M}) = c + \sum\limits_{j = {\ceil{p+1}}}^J D_j(\nu, \mathcal{M}) + (1 - \beta)D_{\ceil{p}}(\nu, \mathcal{M}),
    \label{eq:errorFunction}
\end{align}
where the DWT is computed from the logarithmic data, $\log z(\bm\nu) = ( \log z(\nu_1), \dots, \log z(\nu_K))^T$.
%
%
%

We optimize the parameter $p$ and model $\mathcal{M}$ as follows.
The optimal parameter $\hat{p}$ and model $\hat{\mathcal{M}}$ for additive Raman spectra are ones which minimize
\begin{equation}
    (\hat{p}, \hat{\mathcal{M}}) = \underset{(p, \mathcal{M})}{\arg \min} \left\Vert f(\bm\nu) - \min f(\bm\nu) \right\Vert_1 = \underset{(p, \mathcal{M})}{\arg \min} \sum\limits_{k = 1}^K \left\vert f(\nu_k) - \min f(\bm\nu) \right\vert , \quad p < p_{\rm{max}}
\end{equation}
where we model the constant offset of the background, $c$, such that the minimum value of the spectrum is zero, $f(\bm\nu) - \min f(\bm\nu)$, $p_{\rm{max}}$ is a chosen maximum bound for $p$, and $\left\Vert \, \cdot \, \right\Vert_1$ denotes the $L^1$ norm.
For CARS spectra, we employ a similar scheme:
\begin{equation}
    (\hat{p}, \hat{\mathcal{M}}) = \underset{(p, \mathcal{M})}{\arg \min} \left\Vert \operatorname{Im}\left\{\chi_3(\bm\nu) \right\} - \min \operatorname{Im}\left\{\chi_3(\bm\nu) \right\}\right\Vert_1, \quad p < p_{\rm{max}}
\end{equation}
where we can set $c = 0$, see for example \cite{Kan:16}, and $\operatorname{Im}\left\{\chi_3(\bm\nu) \right\}$ is the Raman spectrum extracted from the corrected spectrum $S(\bm\nu)$.
For details on the extraction of $\operatorname{Im}\left\{\chi_3(\bm\nu) \right\}$, see \cite{Vartiainen:1992, Vartiainen:2006, Liu:2009, Cicerone:2012}.
\section{Experimental details}
\label{sec:experimental}
In the following Sections, we describe the experimental samples used for the numerical examples.
\section*{Raman samples}
The Raman spectra were obtained from a free online database of Raman spectra of pigments used in
modern and contemporary art (The standard Pigments Checker v.5) \cite{ramanDatabase}.
All the spectra used here were corrupted by fluorescence interferences resulting in relatively large background components to the Raman lineshapes.
\section*{CARS samples}
The sugar samples used in the multiplex CARS spectroscopy were equimolar aqueous solutions of D-fructose, D-glucose, and their disaccharide combination, sucrose ({$\alpha$}-D-glucopyranosyl-(1{$\to$}2)-{$\beta$}-D-fructofuranoside).
For sample preparation, the sugar samples were dissolved in buffer solutions (50 mM HEPES, pH=7) at equal molar concentrations of 500 mM \cite{Muller:2007}.
The adenosine phosphate sample was an equimolar mixture of adenosine mono-, di-, and triphosphate (AMP, ADP, ATP) in water in water for a total concentration of 500 mM \cite{Vartiainen:06}.
The adenine ring vibrations \cite{Mathlouthi:1980} are found at identical frequencies for either for AMP, ADP or ATP around 1350 cm$^{-1}$.
The phosphate vibrations between 900 and 1100 cm$^{-1}$ can be used to discriminate between the different nucleotides \cite{Rinia:2006}.
The tri-phosphate group of ATP shows a strong resonance at 1123 cm$^{-1}$, whereas the monophosphate resonance of AMP is found at 979 cm$^{-1}$.
For ADP a broadened resonance is found in between at 1100 cm$^{-1}$.
\section*{Multiplex CARS Spectroscopy}

All CARS spectra used to validate our methodology were recorded using a multiplex CARS spectrometer, the detailed description of which can be found elsewhere \cite{Muller:2002, Rinia:2006}.
In brief, a 10-ps and an 80-fs mode-locked Ti:sapphire lasers were electronically synchronized and used to provide the narrowband pump/probe and broadband Stokes laser pulses in the multiplex CARS process.
The center wavelengths of the pump/probe and Stokes pulses were 710 nm.
The Stokes laser was tunable between 750 and 950 nm.
The sugar spectra were probed within a wavenumber range from 700 to 1250 cm$^{-1}$, and the AMP/ADP/ATP spectrum within a range from 900 to 1700 cm$^{-1}$.
The linear and parallel polarized pump/probe and Stokes beams were made collinear and focused with an achromatic lens into a tandem cuvette.
The latter could be translated perpendicular to the optical axis to perform measurements in either of its two compartments, providing a multiplex CARS spectrum of the sample and of a non-resonant reference under near-identical experimental conditions.
Typical average powers used at the sample were 95 mW (75 mW, in case of AMT/ADP/ATP) and 25 mW (105 mW) for the pump/probe and Stokes laser, respectively.
The anti-Stokes signal was collected and collimated by a second achromatic lens in the forward-scattering geometry, spectrally filtered by short-pass and notch filters, and focused into a spectrometer equipped with a CCD camera.
The acquisition time per CARS spectrum was 200 ms for sugar spectra and 800 ms for the AMP/ADP/ATP spectrum.
\section{Results and discussion}
\label{sec:results}
\begin{figure}[!ht]
    \centering\includegraphics[width=\textwidth]{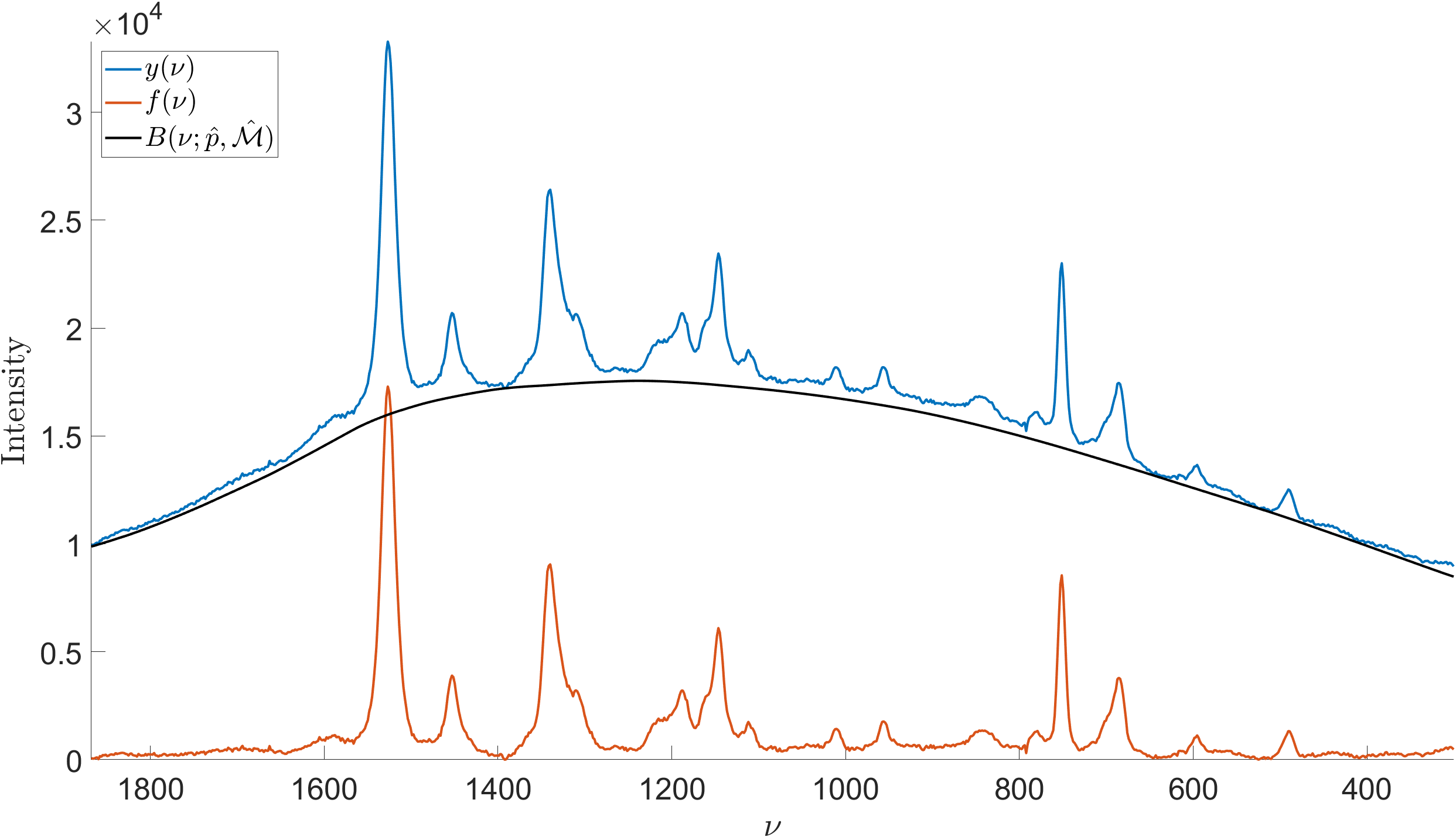}
    \caption{In blue, data for phthalocyanine blue. In red, corrected spectrum $f(\nu)$ with the estimated additive background shown in black. The resulting background was obtained using $ \hat{\mathcal{M}} =$ symlet 5 wavelet with $\hat{p} = 10.663$.}
    \label{im:halo_result}
\end{figure}

\begin{figure}[h!]
    \centering\includegraphics[width=\textwidth]{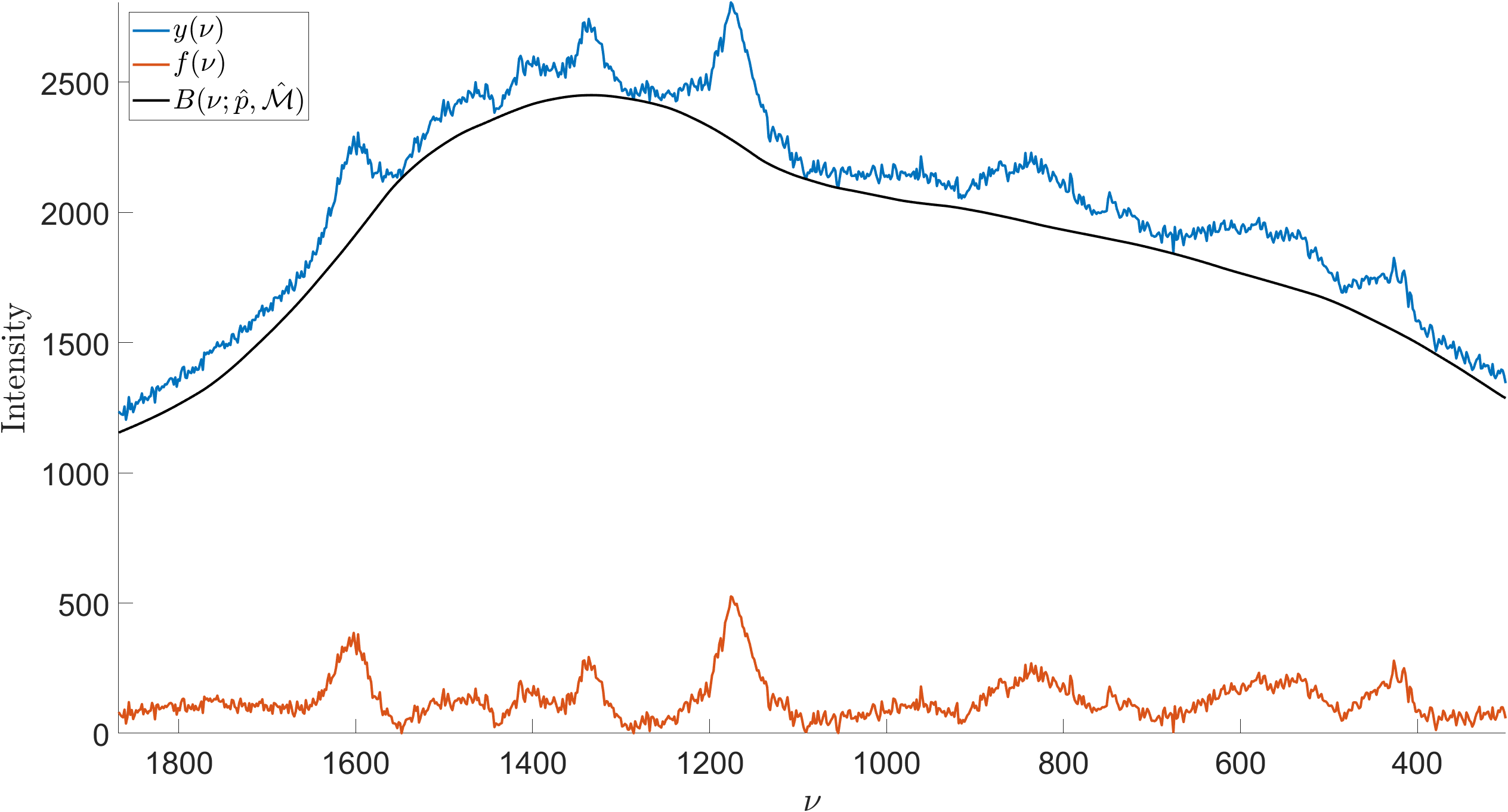}
    \caption{In blue, data for aniline black. In red, corrected spectrum $f(\nu)$ with the estimated additive background shown in black. The resulting background was obtained using $ \hat{\mathcal{M}} =$ symlet 5 wavelet with $\hat{p} = 11.882$.}
    \label{im:pbk_result}
\end{figure}

\begin{figure}[!ht]
    \centering\includegraphics[width=\textwidth]{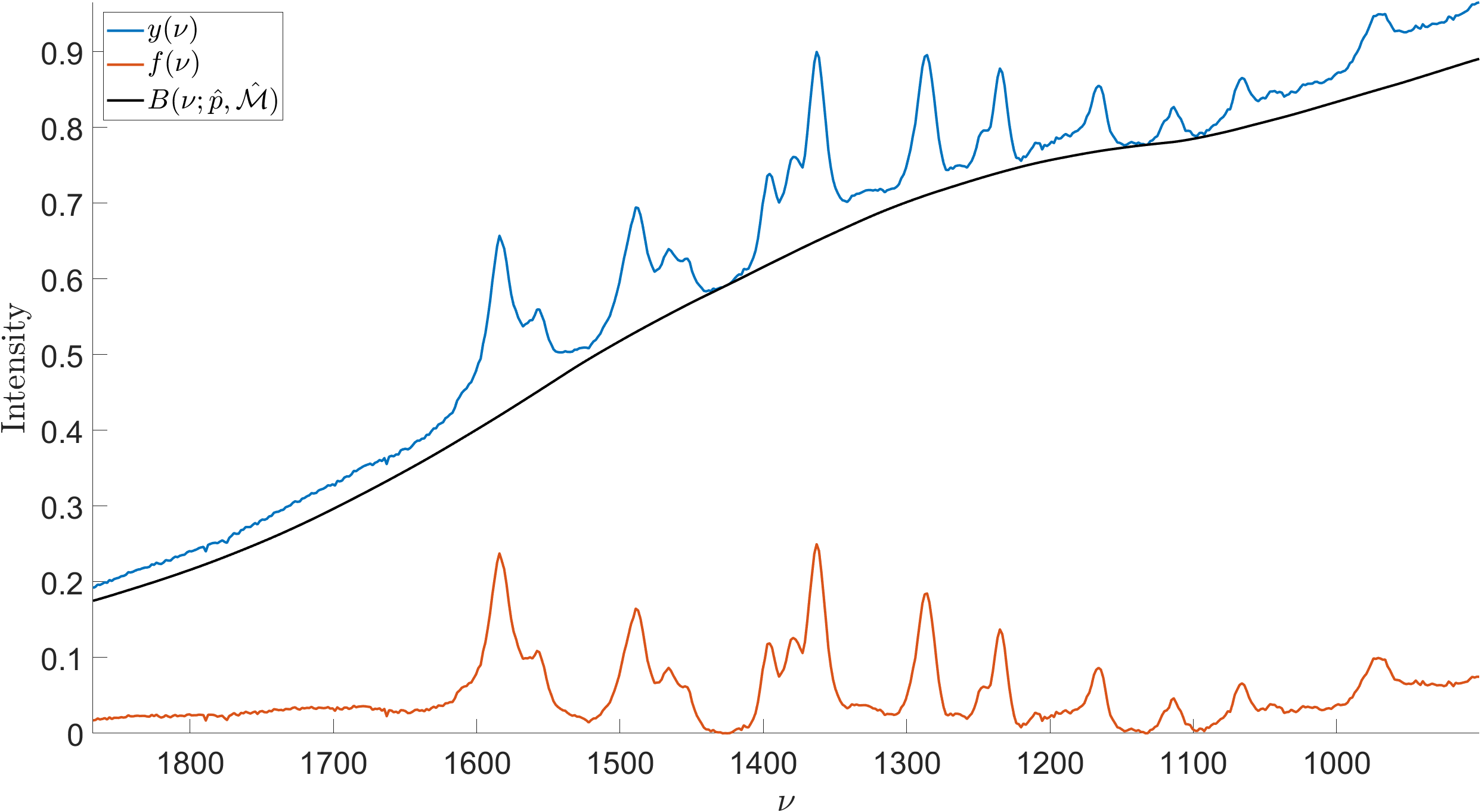}
    \caption{In blue, data for naphthol red. In red, corrected spectrum $f(\nu)$ with the estimated additive background shown in black. The resulting background was obtained using $ \hat{\mathcal{M}} =$ symlet 5 wavelet with $\hat{p} = 11.591$.}
    \label{im:naphtol_result}
\end{figure}

\begin{figure}[h!]
    \centering\includegraphics[width=\textwidth]{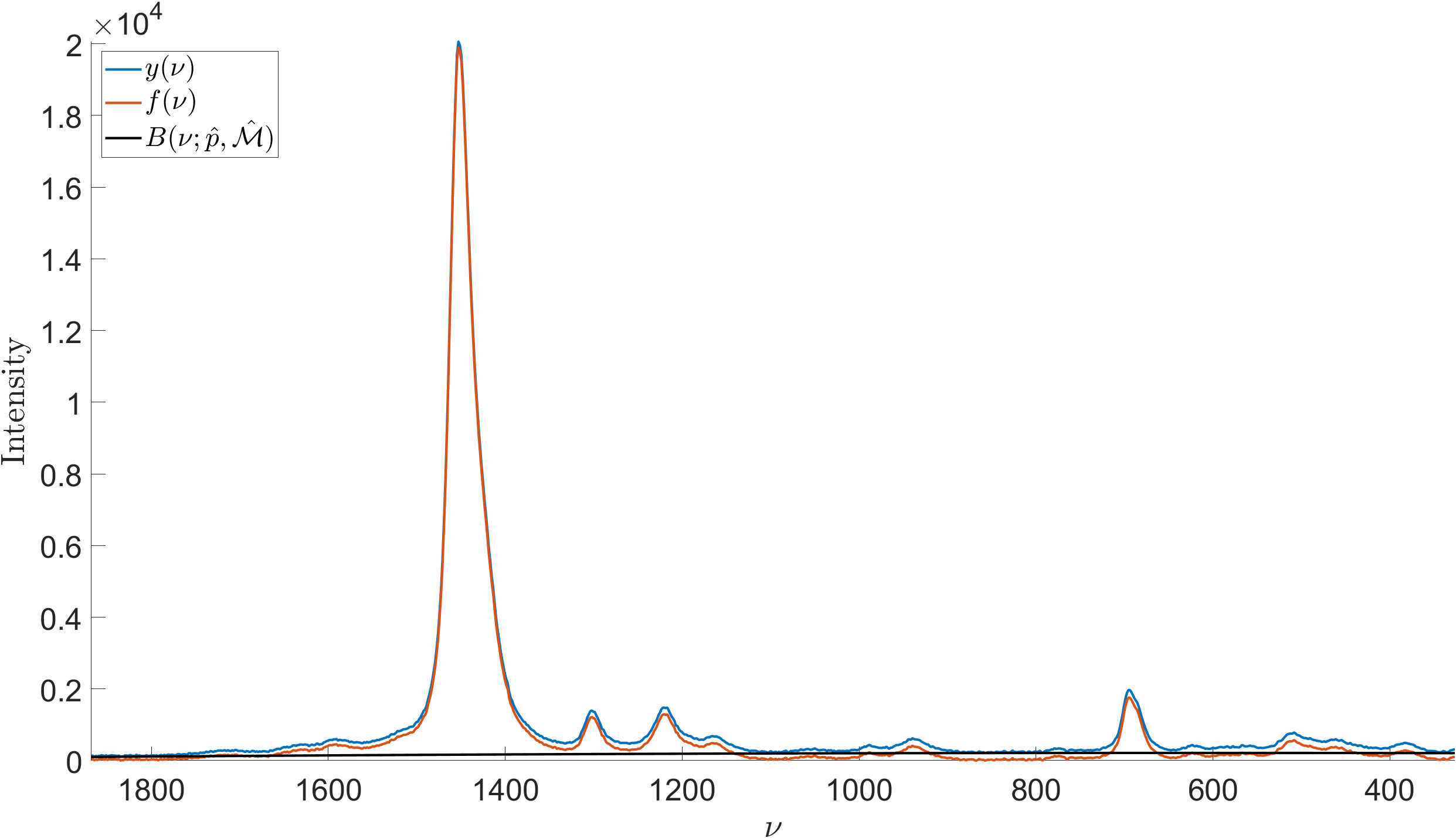}
    \caption{In blue, data for pigment yellow 150. In red, corrected spectrum $f(\nu)$ with the estimated additive background shown in black. The resulting background was obtained using $ \hat{\mathcal{M}} =$ Daubechies 22 wavelet with $\hat{p} = 7.756$.}
    \label{im:nickel_result}
\end{figure}

\begin{figure}[h!]
    \centering\includegraphics[width=\textwidth]{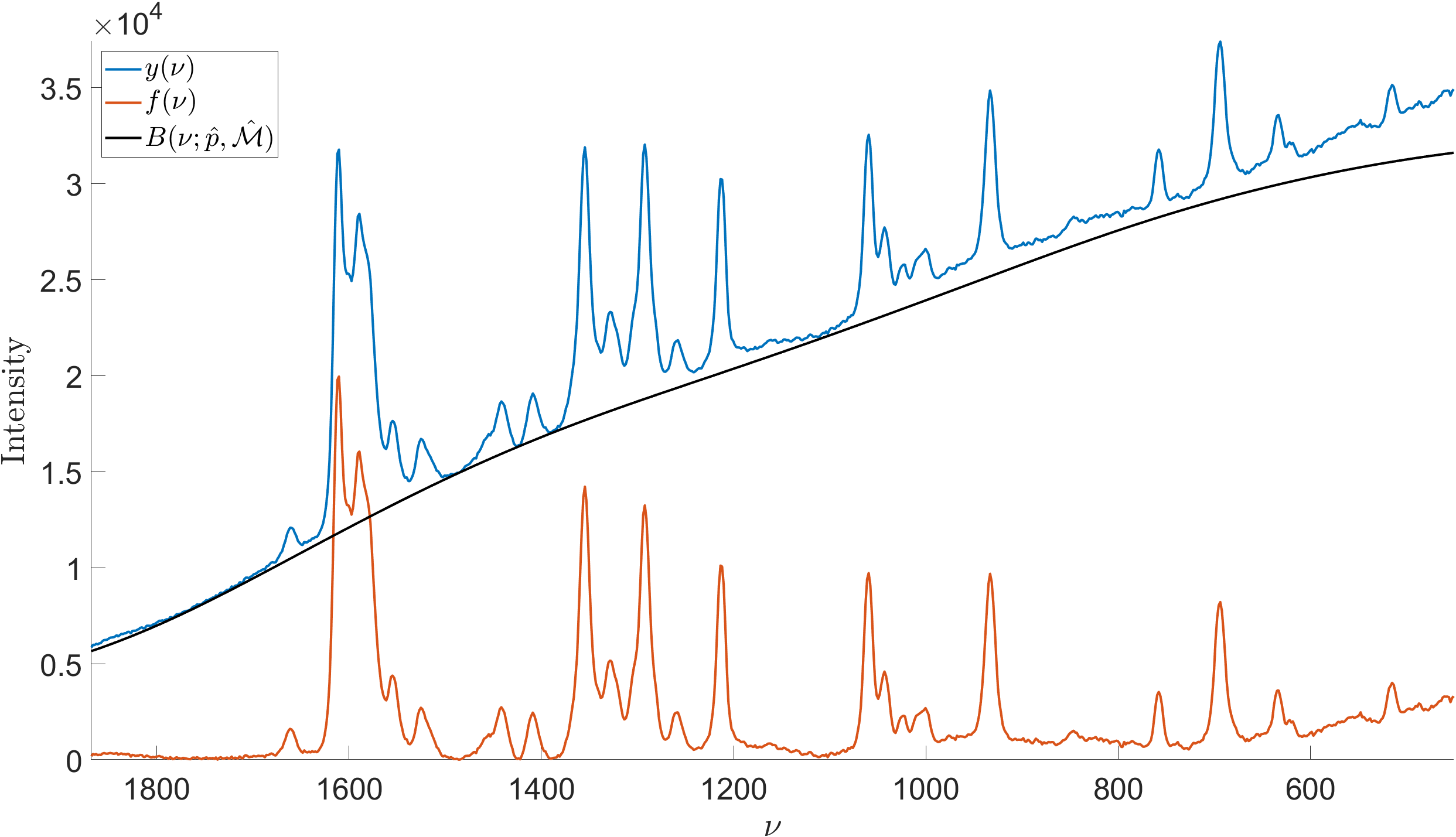}
    \caption{In blue, data for pigment red 264. In red, corrected spectrum $f(\nu)$ with the estimated additive background shown in black. The resulting background was obtained using $ \hat{\mathcal{M}} =$ symlet 11 wavelet with $\hat{p} = 10.555$.}
    \label{im:pyrole_result}
\end{figure}

\begin{figure}[!ht]
    \centering\includegraphics[width=\textwidth]{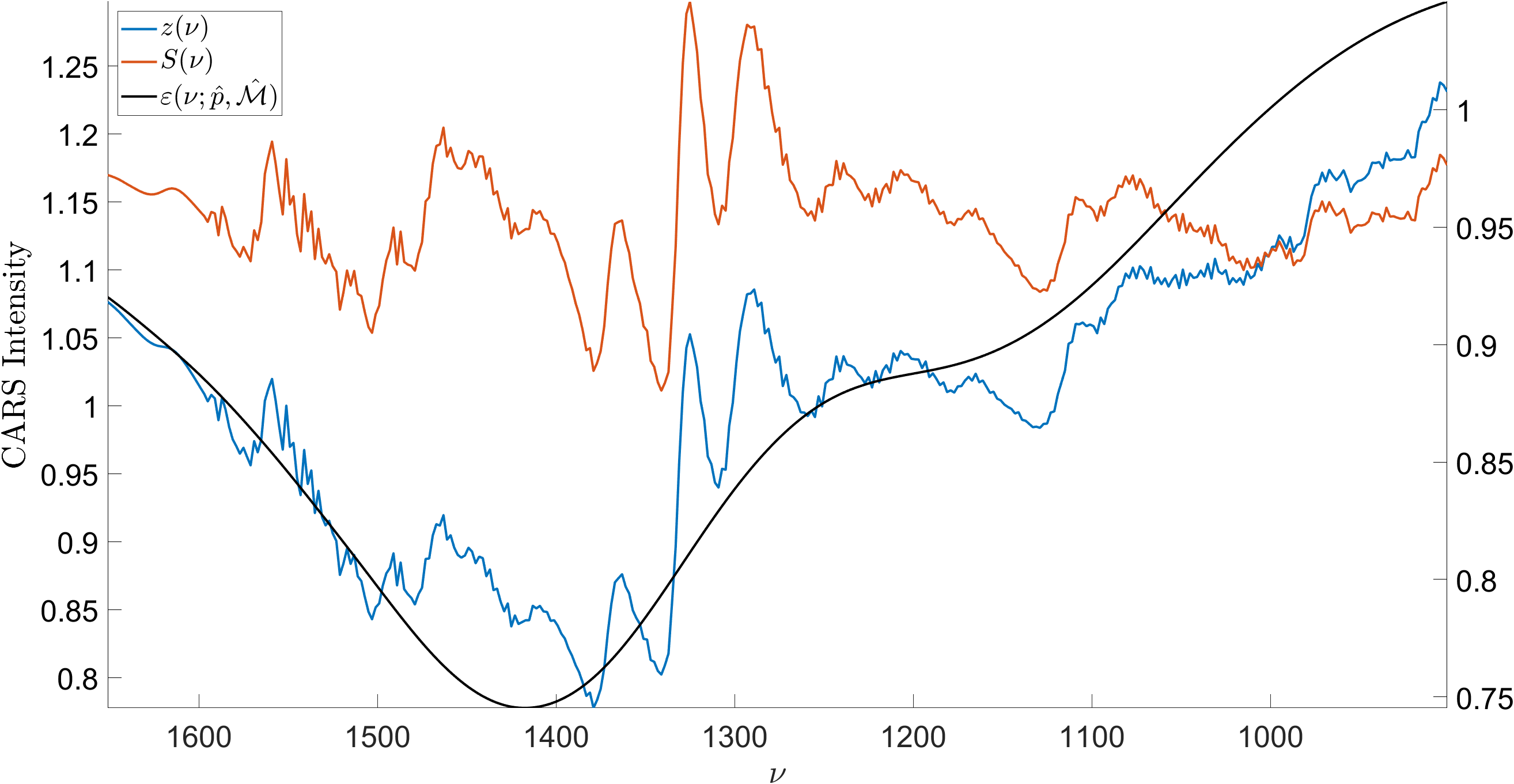}
    \caption{In blue, a measured CARS spectrum for adenosine phosphate. In red, corrected spectrum $S(\nu)$ with the estimated multiplicative background shown in black. The resulting background was obtained using $ \hat{\mathcal{M}} =$ Daubechies 10 wavelet with $\hat{p} = 12.652$.}
    \label{im:cars_amdc_result}
\end{figure}

\begin{figure}[h!]
    \centering\includegraphics[width=\textwidth]{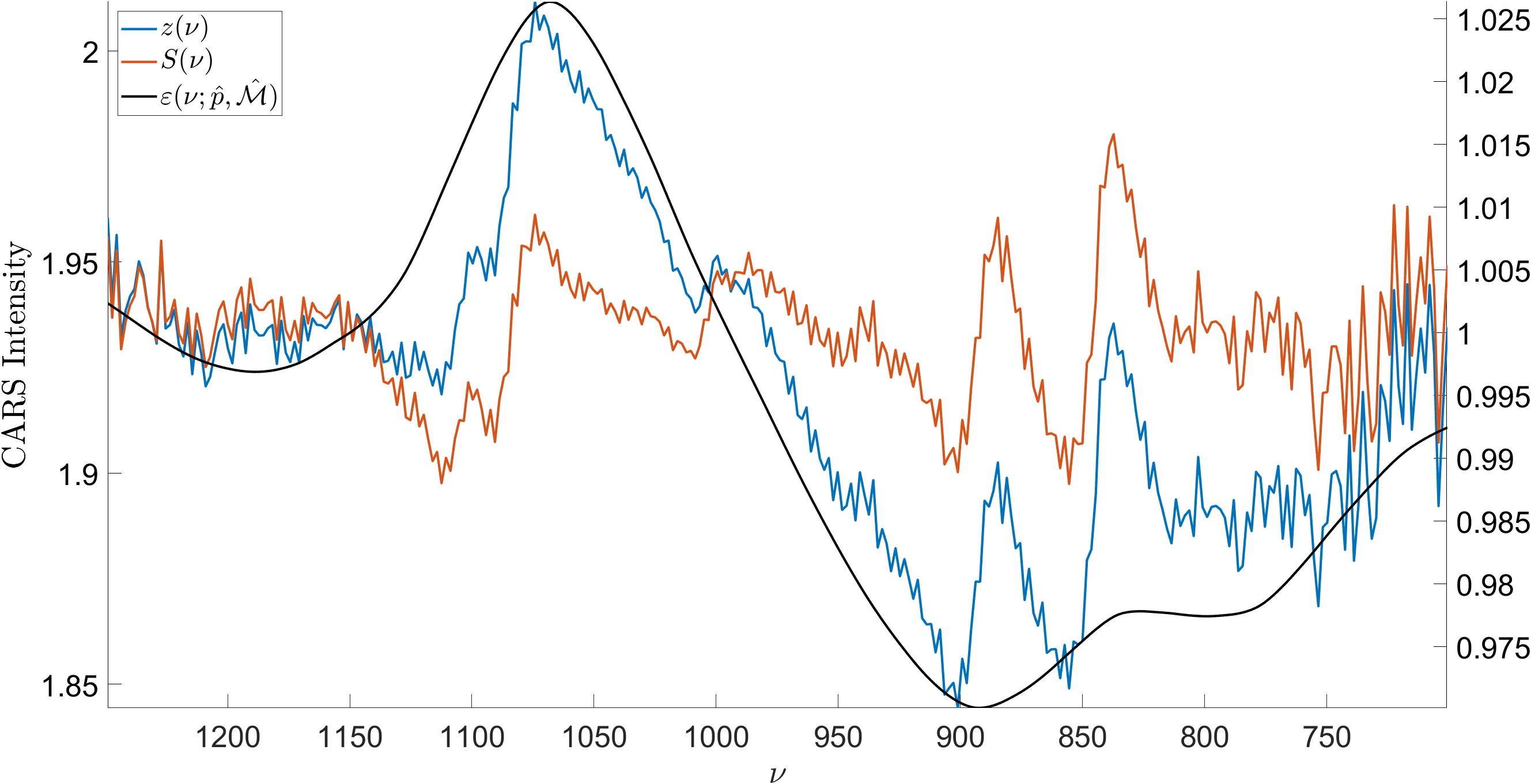}
    \caption{In blue, a measured CARS spectrum for fructose. In red, corrected spectrum $S(\nu)$ with the estimated multiplicative background shown in black. The resulting background was obtained using $ \hat{\mathcal{M}} =$ coiflet 3 wavelet with $\hat{p} = 13.402$.}
    \label{im:cars_fructose_result}
\end{figure}

\begin{figure}[!ht]
    \centering\includegraphics[width=\textwidth]{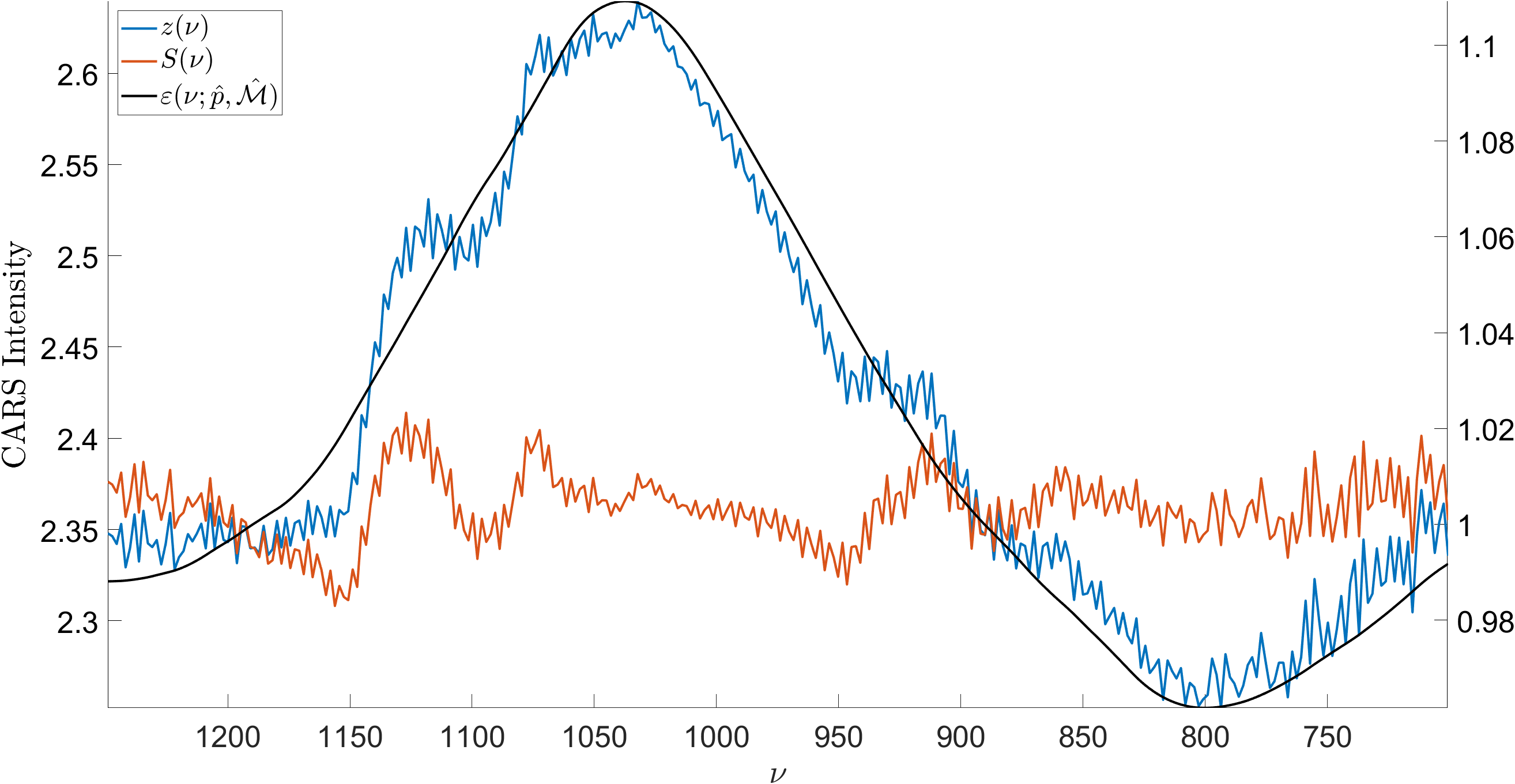}
    \caption{In blue, a measured CARS spectrum for glucose. In red, corrected spectrum $S(\nu)$ with the estimated multiplicative background shown in black. The resulting background was obtained using $ \hat{\mathcal{M}} =$ Daubechies 5 wavelet with $\hat{p} = 13.616$.}
    \label{im:cars_glucose_result}
\end{figure}

\begin{figure}[h!]
    \centering\includegraphics[width=\textwidth]{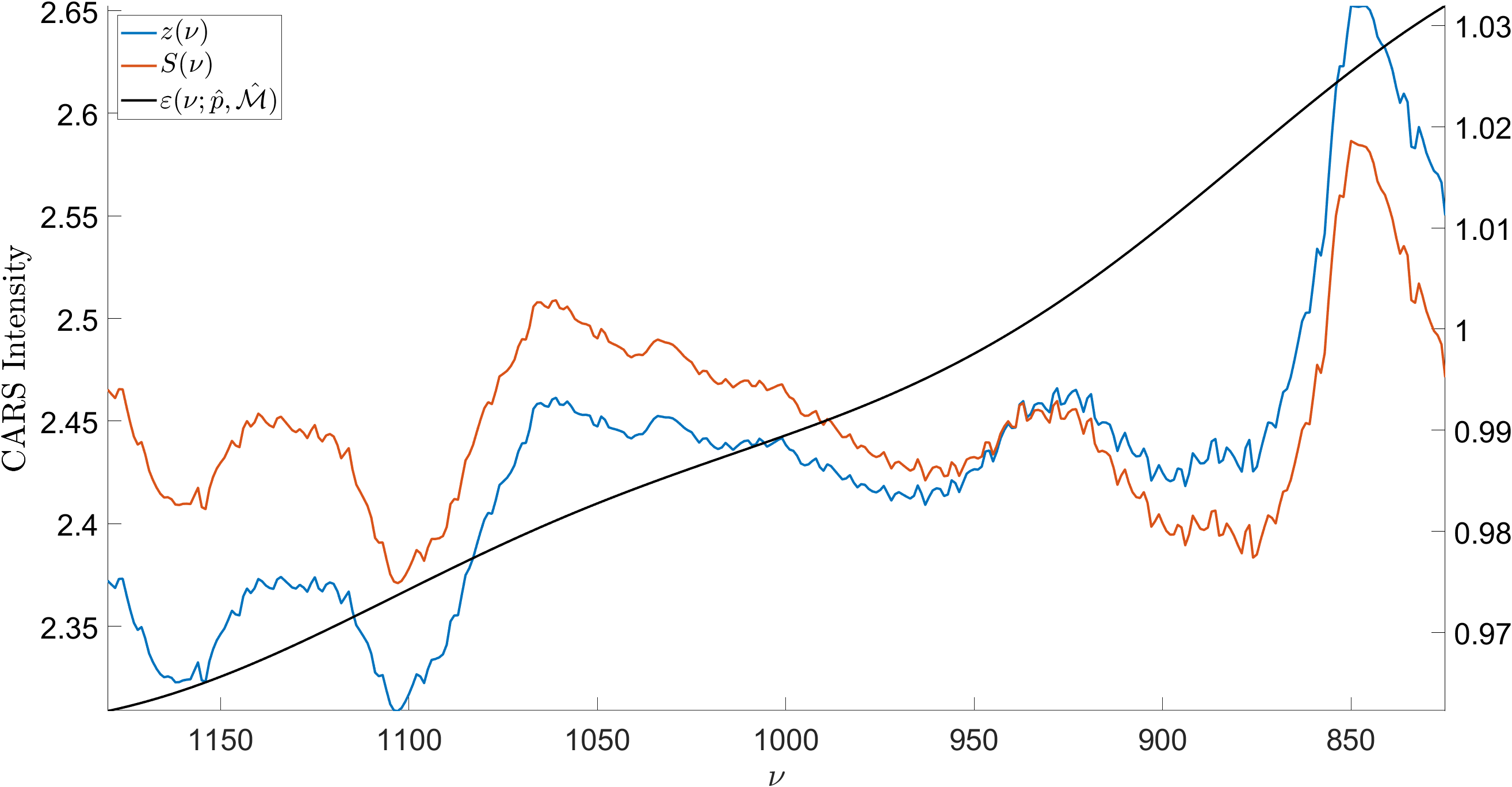}
    \caption{In blue, a measured CARS spectrum for sucrose. In red, corrected spectrum $S(\nu)$ with the estimated multiplicative background shown in black. The resulting background was obtained using $ \hat{\mathcal{M}} =$ Daubechies 35 wavelet with $\hat{p} = 11.702$.}
    \label{im:cars_sucrose_result}
\end{figure}

We apply the method to experimental Raman spectra of phthalocyanine blue, aniline black, and naphthol red for additive background removal and to experimental CARS spectra of adenosine phosphate, fructose, glucose, and sucrose for multiplicative background correction.
The results are presented in Figs. \ref{im:halo_result}--\ref{im:cars_sucrose_result}, respectively, showing the experimental measurements, the estimated additive or multiplicative background, and the corrected spectrum.

For the wavelet models $\mathcal{M}$, Daubechies 5--35, symlet 5--25, and Coiflet 1--5 wavelets were used.
This totals 57 different wavelet bases.

Results for the phthalocyanine blue, aniline black, and naphthol red show smooth backgrounds for the Raman spectra, albeit with some boundary effects for the naphthol red background.
Similar boundary effects are visible in the results for pigment red 264 in Fig. \ref{im:pyrole_result}.
Fig. \ref{im:nickel_result} shows the background removal for the Raman spectrum for pigment yellow 150.
The small and practically constant background is successfully removed and no overfitting is observable.
The optimized wavelet bases were symlet 5, symlet, symlet
5, Daubechies 22, symlet 11 for the phthalocyanine blue, aniline black, naphthol red, pigment yellow 150, and pigment red 264 Raman spectra.

For the CARS spectra, the results are similar to the results presented in \cite{Harkonen:2020}.
The results for the adenosine phosphate, fructose, and glucose CARS spectrum in Figs. \ref{im:cars_amdc_result}--\ref{im:cars_glucose_result} show different slowly varying background signals with roughly similar frequency contents which can also be approximately inferred from the obtained wavelet interpolation parameters which are of similar magnitude, $\hat{p} = 12.652, 13.402, 13.616$, respectively.
The sucrose CARS results in Fig. \ref{im:cars_sucrose_result} show a more linear background which is also evident in the lower wavelet interpolation parameter value, $\hat{p} = 11.702$, implying lower frequency components in comparison to the other three cases.
The optimized wavelet basises for the adenosine phosphate, fructose, glucose and sucrose CARS spectra were Daubechies 38, symlet 12,  symlet 14, and Daubechies 31.

The wavelet background model shows flexible behavior.
For the phthalocyanine blue Raman sample in Fig. \ref{im:halo_result}, the estimated background is practically constant.
In Figs. \ref{im:pyrole_result} and \ref{im:cars_sucrose_result} for the pigment red 264 and sucrose, the behaviour resembles a linear fit.
With phthalocyanine blue, aniline black, and naphthol red Raman samples, as well as, the adenosine phosphate, fructose, glucose CARS samples, the model is able to capture varying degrees of oscillating features.
\clearpage
\section*{Raman}
\vfill
\clearpage
\vfill
\clearpage
\vfill
\clearpage
\section*{CARS}
\vfill
\clearpage
\clearpage
\vfill
\clearpage
\section{Conclusions}
\label{sec:conclusions}
In this study, we proposed an interpolated wavelet-based background model as a tool for modeling background functions for spectra.
We successfully optimize the background parameter and the wavelet basis using an optimization scheme based on minimizing areas of the corrected Raman spectra for both Raman and CARS measurements.
The model can model a flexible family of background functions while being parameterized according with only two parameters for Raman spectra with additive backgrounds and one parameter for CARS spectra with multiplicative backgrounds.
We apply the proposed method to experimental Raman spectra of aniline black and phthalocyanine blue and to experimental CARS spectra of adenosine phosphate, fructose, glucose, and sucrose.
Results for CARS spectra agree with previously obtained results.

\begin{backmatter}
\bmsection{Funding}
The authors were supported by Academy of Finland (grant number 327734).
\bmsection{Disclosures}
The authors declare no conflicts of interest.
\bmsection{Data Availability}
Data and software underlying the results presented in this paper are publicly available in the corresponding author's personal repository on GitHub or from the authors upon reasonable request.
\end{backmatter}
\bibliography{ref.bib}

\end{document}